\documentclass[fleqn,copyright,creativecommons]{eptcs}
\providecommand{\event}{PLACES 2025} 

\usepackage{iftex}
\usepackage[english]{babel}
\usepackage{multicol}
\usepackage{graphicx}
\usepackage{syntax}
\usepackage{amsfonts}
\usepackage{amsmath}

\usepackage[colorinlistoftodos,prependcaption,textsize=normalsize]{todonotes}

\ifpdf
  \usepackage{underscore}         
  \usepackage[T1]{fontenc}        
\else
  \usepackage{breakurl}           
\fi

\title{Static Communication Analysis for Hardware Design}
\author{Mads Rosendahl
\qquad\qquad
Maja H.\ Kirkeby
\institute{Roskilde University\\Computer Science\\Denmark\\
\email{madsr@ruc.dk, kirkebym@acm.org}%
\footnote{This work is supported by the Independent Research Fund Denmark Project no. 2102-00281B}
}}

\newcommand{\titlerunning}{Static Communication Analysis for Hardware Design}
\newcommand{\authorrunning}{Rosendahl \&{} Kirkeby}

\hypersetup{
  bookmarksnumbered,
  pdftitle    = {\titlerunning},
  pdfauthor   = {\authorrunning},
  pdfsubject  = {EPTCS},               
}

\begin{document}
\maketitle

\begin{abstract}
Hardware acceleration of algorithms is an effective method for improving performance in high-demand computational tasks. 
However, developing hardware designs for such acceleration fundamentally differs from software development, as it requires a deep understanding of the highly parallel nature of the hardware architecture. 
In this paper, we present a framework for the static analysis of communication within datapath architectures designed for field-programmable gate arrays (FPGAs). 
Our framework aims to enhance hardware design and optimization by providing insights into communication patterns within the architecture, which are essential for ensuring efficient data handling.

\end{abstract}

\def\opname#1{\mathop{{#1}}\nolimits}
\def\SB#1{\textsf{\textbf{#1}}}
\def\AA{\SB{A}}
\def\RR{\SB{R}}
\def\TT{\SB{T}}
\def\EE{\SB{E}}
\def\DD{\SB{D}}
\def\Nat{{\mathbb{N}}}
\def\Num{{\mathbb{Z}}}
\def\Alnum{{\mathbb{A}}}
\def\II{\opname{\hbox to 5pt{%
  \hbox{\vrule height 1.1pt width 5pt}\kern -3.25pt%
  \raise 1pt\hbox{\vrule height 6pt width 1.5pt}\kern -3pt%
  \raise 7.1pt\hbox{\vrule height 0.9pt width 4.5pt}\kern 0.25pt%
}}}
\def\IIm{\II_m}
\def\IIc{\II_c}
\def\TTm{\TT_m}
\def\TTd{\TT_d}
\def\TTs{\TT_s}
\def\TTt{\TT_t}

\def\lbag{[\![}
\def\rbag{]\!]}
\def\den#1{\lbag#1\rbag}
\def\qt#1{\mbox{\tt '#1'}}
\def\kw#1{\mathrel{\mbox{\bf #1}}}
\let\fun\rightarrow
\def\Vmm{\mbox{\hspace*{ 5mm}}}
\def\Xmm{\mbox{\hspace*{10mm}}}
\let\bot\perp
\def\botd{\,\bot_\Sigma\,}
\let\land\wedge
\let\lor\vee
\def\Reg{\opname{\mbox{Reg}}}
\let \ggl\[\let\ggr\]
\def \[{\ggl\begin{array}{@{}lllll}}
\def \]{\end{array}\ggr}

\def\power{\mathop{\raisebox{1.5pt}{\large$\wp$}}}
\def\Sigmat{\tilde{\Sigma}}
\def\sigmat{\tilde{\sigma}}
\let\lub=\sqcup
\let\glb=\sqcap

\def\Dt{\tilde{D}_\bot}
\def\TTtt{\tilde{\TT}}
\def\RRt{\tilde{\RR}}
\def\EEt{\tilde{\EE}}
\def\SS{\SB{S}}
\def\Iv{\mathop{\sf\bf I\!I}}
\let\union=\cup
\let\cross=\times
\def\widen{\mathop{\raisebox{2pt}{$\bigtriangledown$}}}
\let\empty=\emptyset
\def\Xmm{\hbox to 10mm{\hfil}}
\def\Vmm{\hbox to 5mm{\hfil}}
\let\lub=\sqcup
\let\glb=\sqcap
\def\cnc{\mbox{\tt :}\hspace*{-2pt}\mbox{\tt :}}
\def\pls{${}^{+}$}

\section{Introduction}

Advancements in hardware, coupled with growing awareness of IT energy consumption, have driven interest in reducing energy usage through hardware acceleration of algorithms. In previous work \cite{kirkeby22poster}, we explored and measured the energy savings achieved by transitioning software implementations to hardware designs for Field Programmable Gate Arrays (FPGAs). However, transforming algorithms designed for traditional architectures into highly parallel circuits presents several challenges.

Hardware design remains significantly more complex than software implementation. For instance, algorithms like heapsort can be implemented in software with as little as 30-40 lines of code \cite{rosetta-heapsort}, whereas equivalent hardware descriptions often span 600-700 lines in a hardware description language (HDL) \cite{kirkeby22poster,Yousefifeshki2023}. This stark contrast highlights the complexity of hardware development and the need for better abstraction and analysis tools.

In prior work, we introduced a formal semantics for a subset of the HDL Chisel \cite{Rosendahl25mycroft} and demonstrated its application in bitwidth analysis for optimizing register and wire sizes. 
The bitwidth analysis aims to minimize the number of logical units in hardware designs. 

In this paper, we extend our framework to support static analysis of communication within datapath architectures. This extension provides deeper insight into data movement and helps identify opportunities for improving hardware efficiency. By revealing regions of the design where communication bottlenecks or idle behavior occur, the analysis can guide optimizations such as increasing parallelism or restructuring underutilized components.

Our analysis is based on a subset of the hardware description language Chisel. Chisel \cite{chisel:dac2012} provides a rich set of features for describing hardware at multiple levels of abstraction. By leveraging Chisel, algorithmic solutions can be translated into register-transfer level (RTL) hardware designs, consisting of interconnected registers and logic gates.
Chisel, embedded in the Scala programming language, integrates object-oriented and functional programming paradigms to simplify hardware design. The Chisel framework includes an emulator and a Verilog translator, enabling the realization of designs in hardware.

While HDLs provide fine-grained control over hardware design, high-level synthesis (HLS) tools offer an alternative approach by translating software-like descriptions into hardware specifications. Despite advancements in HLS tools, manually crafted HDL designs often achieve superior efficiency \cite{sozzo2022pushing}. This efficiency gap underscores the importance of algorithmic redesign to fully leverage high-level parallelism.

In this paper, we present a formal semantics for the Core Chisel language, which is the basis for the communication analysis in Section \ref{analysis}.
Our GitHub repository \cite{github-chiselcomm} provides example designs, a translator for converting Core Chisel into standard Chisel, an interpreter based on the formal semantics from Section \ref{semantics}, and an implementation of our analysis from Section \ref{analysis}.

%
%
\section{Core Chisel}
\label{core}

Our subset of Chisel, Core Chisel, focuses on datapath architectures where multiple modules communicate via statically allocated channels. Each module contains registers, data banks, channel connections, and a finite state machine (FSM). Within each FSM state, operations such as register transfers, memory access, and channel communication occur simultaneously.

Evaluation occurs at multiple levels of parallelism: all modules execute concurrently, each state accommodates multiple register transfers, and each register transfer involves evaluating expressions and sub-expressions within the same clock cycle. Despite this parallelism, execution remains deterministic, following a fixed sequence of states for a given initial configuration. This is ensured by the fact that all modules and channels operate within the same hardware circuit and are synchronized by a shared clock.

The communication technique used in datapath architectures is based on data channels
with a ready-valid protocol \cite{dally:vhdl:2016} for synchronization. Chisel \cite{chisel-manual} contains support for port constructors with this protocol. 
Data channels function as one-way connections, serving as outbound streams from one module and inbound streams to another. These channels are statically allocated and are synchronized. Each channel consists of three components: data storage ($\emph{data}$) for writing and reading between modules, along with two boolean values  ($\emph{ready}$  and $\emph{valid}$), indicating readiness for writing and reading operations.

The synchronization allows a one-cycle transition from writing to reading on a channel, followed by an extra cycle to reset the channel for subsequent writing operations. 
Writing is allowed when $\emph{ready}\&!\emph{valid}$ is satisfied. Writing will set $\emph{valid}$ to true. Reading from the channel requires $\emph{ready}\&\emph{valid}$ to be satisfied and will set $\emph{ready}$ to false. 
A reset occurs when $\emph{ready}$ is false, setting $\emph{ready}$ to true and $\emph{valid}$ to false.
After reset, the channel can be used for writing.

Each module operates as a finite state machine (FSM) with a dedicated state register, ensuring synchronized execution of operations.
In order to enter a state involving input/output (I/O) operations, all I/O operations must be feasible, ensuring proper synchronization. 
State operations consist of simple register assignments, reading from and writing to channels, and accessing memory banks. 

Memory access is restricted to at most one read and one write operation per clock cycle to ensure consistency.
A value read from memory becomes available only at the end of the clock cycle. It must be stored in a register before being used in subsequent computations.
Part of the register transfer operations within a clock cycle involves computing a new value for the state register. This also means that the condition in a state is evaluated using the values registers had at the start of the clock cycle and not after any updates happen in the state.

%
%
\subsection{Abstract syntax} \label{sect-abs-syntax}

The abstract syntax for Core Chisel is presented below.
Section \ref{semantics} contains semantics for the language.

A program has three parts: the creation of module instances, the connection of output channels to input channels, and the declaration of the modules. 
Module declarations work similarly to class declarations in object-oriented languages, allowing one or more instances of a module to be created.

Each module contains its own set of registers and memory banks, which are not shared with other modules, as well as channels used for inter-module communication. A module also includes a finite state machine, where each state specifies a set of register transfers and a transition to the next state. All operations within a state are executed simultaneously in a single clock cycle, and multiple modules can operate concurrently. States that involve communication are assumed to include a \verb'when' clause, ensuring that all communication conditions are satisfied before execution. If communication is not possible, the module remains idle until the necessary conditions are met.

\begin{quote}
\setlength{\grammarindent}{7em}
\begin{grammar}
<program> ::= <module>\pls <connnection>* <module-decl>\pls 

<module> ::= `val' <ident> `=' `Module' `(' <ident> `)' 

<connection> ::= <ident> `.' <ident> `<>' <ident> `.' <ident>

<module-decl> ::= `module' <ident> <declaration>* <state>\pls

<declaration> ::= `int'  <ident> [ `=' <number> ] 
\alt  `int'   `['  <number> `]' <ident> 
\alt  `instream'    <ident> 
\alt  `outstream'   <ident> 

<state> ::= `state' <number> [ `when' <expr> ] <statement>* `goto' <gotoexp>

<statement> ::= <ident> `=' <expr> 
\alt <ident> `[' <expr> `]' `=' <expr> 
\alt <ident> `=' <ident> `[' <expr> `]' 
\alt <ident> `.' `write' `(' <expr> `)'
\alt <ident> `=' <ident> `.' `read' `(' `)'

<expr> ::= <ident> 
\alt  <number>
\alt  <expr> <operation> <expr>
\alt  `Mux' `(' <expr> `,' <expr> `,' <expr> `)'
\alt  <ident> `.' `ready'
\alt  <ident> `.' `valid'

<operation> ::= `+'  | `-'  | `*'  |  `/'  | `

<gotoexp> ::= <number> 
\alt  `Mux' `(' <expr> `,' <gotoexp> `,' <gotoexp> `)'

\end{grammar}
\end{quote}

\subsection{Example} \label{ex:sendrec-code}

Below, we have a small example of a hardware description with modules that communicate via a channel.
The sender module will transmit along the channel, and the receiver module will find the maximum number received. The receiver will use an unnecessary number of states in the receiving process so that reading from the channel will not block. In the analysis in section \ref{analysis} we examine the communication patterns in hardware designs.

\begin{multicols}{2}
\begin{quote}\small
\begin{verbatim}
val sender = Module(Sender)
val receiver = Module(Receiver)
sender.out <> receiver.in


module Sender
int i = 0
outstream out
state 1 when out.ready()
  out.write(i)
  i = i + 1
  goto Mux(i < 5,1,2)
state 2
  goto 2







\end{verbatim}
\end{quote}
\begin{quote}\small
\begin{verbatim}
module Receiver
int x = 0
int y = 0
int j = 0
instream in
state 1
  j = 0
  goto 2
state 2
  x = 0
  goto 3
state 3 when in.valid()
  x = in.read()
  goto 4
state 4
  y = Mux(x > y,y,x)
  goto 5
state 5
  j = j + 1
  goto Mux(j < 5,3,6)
state 6
  goto 6
\end{verbatim}
\end{quote}
\end{multicols}

The design includes declarations of the modules, preceded by the creation of module instances and the connection of outstreams from one instance to instreams of another.
The sender module attempts to write to the channel in every clock cycle. In contrast, the receiver performs several operations between each state that includes a read from the channel. As a result, the expected behavior is that the receiver should never have to wait for input, while the sender may experience one or more cycles of delay before it can transmit data to the channel.

While hardware designs can, in principle, run indefinitely, this example reaches a stable configuration in which each module enters a state where no further changes occur to state variables, channels, or registers.

%
%
\section{Semantics}
\label{semantics}
\def\paragraph#1{\bigskip\par\noindent{\bf #1} }
\def\paragraphx#1{\vspace*{0pt plus 2pt}\par\noindent{\bf #1} }

The semantics describes the design as a state transition function where each step corresponds to a clock cycle. The semantics is implemented as an interpreter in our 
GitHub repository \cite{github-chiselcomm} 
and we can validate the semantics by showing that the same steps will be performed by the interpreter and the Chisel emulater when the abstract syntax is translated into normal Chisel syntax.

\paragraph{State environment.}
The state of the hardware description is here represented as a mapping of registers, channels, and memory banks to values.
We consider each module and each channel as a separate namespace. 
Modules are identified by their name, and we give each channel a number to identify it in the environment. For memory banks, we treat each location in the bank as a separate register and in the environment reference the individual elements by a concatenation  ($\cnc$) of the name of the memory bank and the index.  The domain for state environments is the set $\Sigma$.
\[
\sigma \in\Sigma : \Reg \fun  \Num_\bot  \Xmm\Xmm
r\in \Reg : (\Nat \cup \Alnum) \times\Alnum
\]

Updating an environment with values from another environment is written as
\[
\sigma_1 \oplus \sigma_2 = \lambda x .\kw{if} \sigma_1(x)=\botd\; \kw{then} \sigma_2(x) \kw{else} \sigma_1(x)
\]

\paragraph{Initial environment.}
A program's initial state environment consists of initializing all registers and memory banks in modules and connecting channels to streams in modules.

The semantic function $\II$ returns the initial environment for the design in which all registers within module instances are initialized to zero, and each module's state register is set to 1. Channel numbers are assigned to instreams and outstreams within module instances, and both the status bits and data components of the channels are initialized accordingly. A single module declaration can correspond to one or more module instances. To distinguish between instances, the interpretation of a module declaration is parameterized by the module name and the instance name.

\[
\II\den{ mod^+ con^*  mdcl^+} =
\IIc\den{con^*} \oplus \IIm\den{mod^*}(\DD\den{mdcl^*})
\\
\IIc\den{con_1 \ldots con_n} = \IIc\den{con_1}\; 1 \oplus \cdots \IIc\den{con_1}\; nn
\\
\IIc\den{m_1.out_1 \qt{<>} m_2.in_2 }\; i =
[(m_1.out_1) \mapsto i] \oplus [(m_2.in_2) \mapsto i] \oplus
\\ \Xmm [(i,\qt{ready}) \mapsto 0] \oplus
        [(i,\qt{valid}) \mapsto 0] \oplus
        [(i,\qt{data}) \mapsto 0]  \oplus
\\ \Xmm  [(m_1,out_1) \mapsto 0]  \oplus
        [(m_2,in_2) \mapsto 0]  
\\
\IIm\den{mod_1 \cdots mod_n}\; d =
  \IIm\den{mod_1}\; d \oplus\cdots\oplus \IIm\den{mod_n}\; d 
\\
\IIm\den{\qt{val} m \qt{=} \qt{Module}\qt{(} M \qt{)} }d =  d\; M\; m
\]
The $\DD$ function defines the
declaration and initialization of registers and memory banks in modules.
\[
\DD\den{mdcl_1\ldots mdcl_n}\; M\; m =
\DD\den{mdcl_1}\; M\; m \oplus \cdots\oplus \DD\den{mdcl_n}\; M\; m
\\
\DD\den{\qt{module} M_1 decl\, states}\; M\; m = 
  \kw{if} M=M_1 \kw{then} \DD\den{decl} m \kw{else} \botd
\\
\DD\den{d_1 \cdots d_n}\; m = \DD\den{d_1}\;m \oplus \DD\den{d_n}\;m
\oplus [(m,\qt{state})\mapsto 1]
\\
\DD\den{\qt{int} x = n}\; m =[(m,x)\mapsto n]
\\
\DD\den{\qt{int} [n] a }\; m =[(m,a\cnc 0)\mapsto n] \oplus\cdots\oplus [(m,a\cnc (n-1)))\mapsto n] 
\]

\paragraphx{Channel reset.}
Reset of channels in environment $\sigma$ is done in each cycle for channels where the data has been read. Reading data from a channel will set $\qt{ready}$ to 0 and the reset will set it to 1 so the channel 
can be used for writing.

During each clock cycle, the status bits of a channel may be modified by the sender, the receiver, or a centralized reset mechanism. However, since states can only be entered when all communication operations are possible, at most one of these entities can update the status bits in a given cycle. The channel reset operation is combined with the state transition operations of module instances to define the complete set of environment updates that occur within a single clock cycle.

\[
\RR\den{con_1 \cdots con_n}\;\sigma = 
  \RR\den{con_1}\;\sigma\oplus\cdots\oplus\RR\den{con_n}\;\sigma
\\
\RR\den{m_1.out_1 \qt{<>} m_2.in_2 }\;\sigma=
\kw{let} c = \sigma(m_1,out_1) 
\\\Xmm
    \kw{if} \sigma(c,\qt{ready})=0 \kw{then}
    [(c,\qt{ready})\mapsto 1,(c,\qt{valid})\mapsto 0] \kw{else}\botd
\]

\paragraphx{State transitions.}
The semantic functions record changes to the state environment $\sigma$
for each module in each clock cycle. 
The function $\TT_d$ is applied to module declarations, $\TT_t$ to the states in modules and $\TT_s$ to statements in states.

The function $\TT$ defines the changes to the environment that occur in each clock cycle. It is composed of the state transitions from each module instance along with the channel reset operations. As with the construction of the initial environment, the interpretation of a module declaration is parameterized by a name of a corresponding module instance.

For each module instance, the state transition depends on its current state variable and whether the communication required in that state can proceed. Each state also includes an expression that determines the next state. This expression is evaluated in the same environment as the register transfers within the state. At the hardware level, this evaluation occurs within the same clock cycle as the register transfers and is based on register values from before the cycle began.

\[
\TT\den{ mod^+ con^*  mdcl^+} \sigma =
(\TTm\den{mod^*}(\TTd\den{mdcl^*})\;\sigma) \oplus
\RR\den{con^*}\sigma 
\\
\TTm\den{mod_1\cdots mod_n}\; d\;\sigma=
\TTm\den{mod_1}\;d\;\sigma \oplus \cdots\oplus
\TTm\den{mod_n}\;d\;\sigma 
\\
\TTm\den{\qt{val} m \qt{=} \qt{Module}\qt{(} M \qt{)}}\;d\;\sigma\; m =
d \; \sigma\; M\; m
\\
\TTd\den{mdcl_1\cdots mdcl_n}\;\sigma\;M\;m =
\TTd\den{mdcl_1}\;\sigma\;M\;m \oplus \cdots\oplus
\TTd\den{mdcl_n}\;\sigma\;M\;m 
\\
\TTd\den{\qt{module} M_1 decl\, states}\; \sigma\;M\; m = 
\kw{if} M=M_1 \kw{then} \TTt\den{states}\sigma\;m
\kw{else} \botd
\]
\[
\TTt\den{state_1\cdots state_n}\;\sigma\; m =
\TTt\den{state_1}\sigma\; m \oplus\cdots\oplus
\TTt\den{state_n}\sigma\; m
\\
\TTt\den{\qt{state} n\qt{when} e\; s_1 \cdots s_n \qt{goto} e_g}\;\sigma\; m =
\\\Xmm
  \kw{if} \sigma(m,\qt{state}) \neq n \kw{then} \botd\, \kw{else}
\\\Xmm
  \kw{if} \EE\den{e}\;\sigma\; m \neq 1 \kw{then} \botd\, \kw{else}
\\\Xmm
   \TTt\den{s_1}\;\sigma\; m \oplus\cdots\oplus \TTt\den{s_n}\;\sigma\; m  
   \oplus[(m,\qt{state})\mapsto \EE\den{e_g}\;\sigma\; m]
\]
Values of locations in memory banks are stored in the environment under a name obtained by concatenating the name of the memory bank with the index using a concatenation operation ($\cnc$).
\[
\TTs\den{x\qt{=} e}\;\sigma\; m = [(m,x)\mapsto \EE\den{e}\sigma m]
\\
\TTs\den{a[e_1]\qt{=} e_2}\;\sigma\; m = 
  [(m,a\cnc\EE\den{e_1}\;\sigma\; m)\mapsto \EE\den{e_2}\;\sigma\; m]
\\
\TTs\den{x\qt{=} a[e_1]}\;\sigma\; m = 
  [(m,x)\mapsto \sigma(m,a\cnc\EE\den{e_1}\;\sigma\; m)]
\\
\TTs\den{out\qt{.write(} e\qt{)}}\;\sigma\; m =
\\\Xmm
[(\sigma(m,out),\qt{data}) \mapsto \EE\den{e}\sigma\; m,
(\sigma(m,out),\qt{valid}) \mapsto 1]
\\
\TTs\den{x \qt{=}in\qt{.read()}}\;\sigma\; m =
\\\Xmm
[(m,x) \mapsto \sigma(\sigma(m,in),\qt{data}),
(\sigma(m,in),\qt{ready}) \mapsto 0]
\]
%
%
When we read and write to channels, we change the status bits in the channel, and before reading or writing, we check whether the channel is ready for reading or writing.
\[
\EE\den{n}\;\sigma\; m = n
\\
\EE\den{v}\;\sigma\; m = \sigma(m,v)
\\
\EE\den{\qt{Mux(} e_1,e_2,e_3\qt{)}}\;\sigma\; m =
\kw{if} \EE\den{e_1}\;\sigma\; m =1 \kw{then} \EE\den{e_2}\;\sigma\; m
\kw{then} \EE\den{e_3}\;\sigma\; m
\\
\EE\den{ e_1 \;op\; e_2}\;\sigma\; m =
op( \EE\den{e_1}\;\sigma\; m, \EE\den{e_2}\;\sigma\; m)
\\
\EE\den{in\qt{.ready()}}\;\sigma\; m = 
  \sigma(\sigma(m,in),\qt{ready})=1 \land
  \sigma(\sigma(m,in),\qt{valid})=0 
\\
\EE\den{out\qt{.valid()}}\sigma\; m = 
  \sigma(\sigma(m,out),\qt{ready})=1 \land
  \sigma(\sigma(m,out),\qt{valid})=1 
\]

\paragraph{Initial to final environment.}
The transition function $\TT$ collects changes to the environment in a step.
The interpretation is then repeated transitions until stability. 
\[\kw{Interpreter}\!\den{prg} =
(\kw{fix} \lambda F \lambda  \sigma . F(\TT\den{prg} \sigma \oplus \sigma)) \sigma_0
\Xmm \sigma_0 = \II\den{ mod^+ con^*  mdcl^+}
\]

The interpretation of a design maps the initial environment to a final environment—one in which no further changes are produced by the state transition function. Our GitHub repository \cite{github-chiselcomm} provides an implementation of this semantics, along with a translator from the abstract syntax to Chisel code. This allows the interpreter to be compared directly with the behavior of the Chisel emulator.

%
%
\section{Communication Analysis}
\label{analysis}

The communication analysis follows the structure of 
the standard semantics, but can be seen as an abstraction where we record sets of possible values for channel status bits, but ignore values in registers and memory banks.
We distinguish between four kinds of registers:
$ \Reg_d = \Nat \times\{\kw{data}\}$ is data channels, $  \Reg_{rv} =\Nat\times\{\kw{ready},\kw{valid}\} $
are status wires in channels, $\Reg_r$ are ordinary registers and memory banks in modules, and $\Reg_s$ are state number and channel stream numbers in modules.
%
\[
\sigmat \in\Sigmat : ( \Reg_{rv} \cup \Reg_s ) \fun (\power(2) \cup \Nat)
\]
where $\power(2)$ are the subsets of $\{1,2\}$.

The state transition is a mapping of an abstract environment  ($\sigmat$) into a set of possible next environments
($\power(\Sigmat)$). 
The aim is to find the set of possible abstract environments reachable from the initial environment. This is similar to the construction of minimal function graphs for a first-order functional language  \cite{jones1986data}.

\paragraph{Abstraction.}
The abstraction of environment in the standard interpretation ignores data registers and has status bit values as singleton sets.
\[
\alpha : \Sigma \fun \Sigmat
\\
\alpha( \sigma) = \lambda x .
\kw{if} x\in\Reg_{rv}  \kw{then} \{\sigma(x)\} \kw{else if} x \in\Reg_s \kw{then} \sigma(x) \kw{else} \bot
\]

As initial environment we just use the abstraction of the initial environment in the standard interpretation.
\[
\sigmat_0 = \alpha(\II\den{ mod^+ con^*  mdcl^+}  )
\]

\paragraphx{Collecting environments.} The analysis is done by iterating a set of abstract environments until stability. As for minimal function graphs it will include all abstractions of all intermediate states.
\[
\TTtt\den{prg} : \power(\Sigmat)
\\
\TTtt(prg) = \kw{fix}\lambda \theta .  \{\sigmat_0\} \cup
\{ \TTtt_m\den{prg} \sigmat \oplus \RR\den{prg}\sigmat \oplus \sigmat \mid  \sigmat \in \theta \}
\]

\paragraph{State transitions.}
We will here assume we have a numbering of all module instances $m_1,\ldots,m_n$ and corresponding module declarations $Mod_1,\ldots, Mod_n$. The possible state transitions are then the set of all combinations of state transitions for each module.
\[
\TTtt_m\den{m_1\cdots m_n} \; \sigmat  =
\TTtt_t\den{Mod_1} \sigmat m_1 \otimes \cdots  \otimes \TTtt_t\den{Mod_n} \sigmat m_n
\]
%
%
%
\[
\TTtt_t\den{s}\; \Sigmat \fun \Alnum \fun  \power(\Sigmat)
\\
\TTtt_t\den{state_1\cdots state_n}\;\sigmat\; m =
\TTtt_t\den{state_1}\sigmat\; m \oplus\cdots\oplus
\TTtt_s\den{state_n}\sigmat\; m
\\
\TTtt_t\den{\qt{state} n\qt{when} e\; s_1 \cdots s_n \qt{goto} e_g}\;\sigmat\; m =
\\\Xmm
  \kw{if} \sigma(m,\qt{state}) \neq n \kw{then} \botd\, \kw{else}
\\\Xmm
  \kw{if} \EEt\den{e}\;\sigmat\; m \neq 1 \kw{then} \botd\, \kw{else}
\\\Xmm
   \TTtt_s\den{s_1}\;\sigmat\; m \oplus\cdots\oplus \TTtt_s\den{s_n}\;\sigma\; m  
   \oplus[(m,\qt{state}\mapsto \EEt\den{e_g}\;\sigma\; m]
\]
\[
\TTtt_s\den{s}\; \Sigmat \fun \Alnum \fun  \Sigmat
\\
\TTtt_s\den{s_1 \cdots s_n}\;\sigmat\; m  
&= \TTtt_s\den{s_1}\;\sigmat\; m \oplus \cdots \TTtt_s\den{s_1}\;\sigmat\; m 
\\
\TTtt_s\den{out\qt{.write(} e\qt{)}}\;\sigmat\; m 
&= 
[ (\sigma(m,out),\qt{valid}) \mapsto \{1\}]
\\
\TTtt_s\den{x \qt{=}in\qt{.read()}}\;\sigmat\; m 
&= 
[(\sigma(m,in),\qt{ready}) \mapsto \{0\}]
\]


\paragraph{Channel reset and expressions.}
\[
\RRt\den{con_1 \cdots con_n}\;\sigmat = 
  \RRt\den{con_1}\;\sigmat\oplus\cdots\oplus\RRt\den{con_n}\;\sigmat
\\
\RRt\den{m_1.out_1 \qt{<>} m_2.in_2 }\;\sigmat=
\kw{let} c = \sigmat(m_1,out_1) 
\\\;\;
    \kw{if} \sigmat(c,\qt{ready})=\{0,1\} \kw{then}
    [(c,\qt{ready})\mapsto \{1\},(c,\qt{valid})\mapsto \{0\}\cup\sigmat(c,\qt{valid})] 
\\\;\;
    \kw{else if} \sigmat(c,\qt{ready})=\{0\} \kw{then}
    [(c,\qt{ready})\mapsto \{1\},(c,\qt{valid})\mapsto \{0\}] \kw{else}\botd 
\]
\[
\EEt_g  \den{n}\;  = \{n\}
\\
\EEt_g  \den{\qt{Mux(} e_1,e_2,e_3\qt{)}}\;  = \EEt_g e_1 \cup \EEt_g e_2
\]
\[
\EEt\den{in\qt{.ready()}}\;\sigmat\; m =& 
 \{1\mid 1\in\sigma(\sigma(m,in),\qt{ready}) \land
  0\in\sigma(\sigma(m,in),\qt{valid}) \}\cup
\\&
 \{0\mid 1\not\in\sigma(\sigma(m,in),\qt{ready}) \lor
  0\not\in\sigma(\sigma(m,in),\qt{valid}) \}
\\
\EEt\den{in\qt{.valid()}}\sigmat\; m = &
 \{1\mid 1\in\sigma(\sigma(m,in),\qt{ready}) \land
  1\in\sigma(\sigma(m,in),\qt{valid}) \}\cup
\\&
 \{0\mid 1\not\in\sigma(\sigma(m,in),\qt{ready}) \lor
  1\not\in\sigma(\sigma(m,in),\qt{valid}) \}
\]

\paragraph{Validity.}
The analysis is implemented and can be found in our GitHub repository. The validity of the analysis
can be examined by testing that the abstract state transition gives a safe approximation of the abstraction of state transitions in the standard interpretation.
\[
\TTtt_m\den{prg} (\alpha(\sigma)) \sqsupseteq \alpha(\TT_m\den{prg} \sigma)
\]

\section{Example}
%

The example in Section \ref{core} illustrates a scenario with a fast sender, capable of writing data in every clock cycle, and a slower receiver that processes data across multiple stages. Analyzing the code reveals that when the receiver is in state 3, the channel either contains valid data—allowing the receiver to proceed without waiting—or remains empty if the sender is in state 2. The analysis successfully tracks the flow of status bits in the communication but does not establish that the number of values sent along the channel matches the number of values expected by the receiver.

\begin{quote}\small
\begin{verbatim}
receiver,state : 1
// 1,ready: 0  1,valid: 0  sender,state: 1  
receiver,state : 2
// 1,ready: 1  1,valid: 0  sender,state: 1    
receiver,state : 3
// 1,ready: 1  1,valid: 1  sender,state: 2    
// 1,ready: 1  1,valid: 0  sender,state: 2    
// 1,ready: 1  1,valid: 1  sender,state: 1    
receiver,state : 4
// 1,ready: 0  1,valid: 1  sender,state: 2    
// 1,ready: 0  1,valid: 1  sender,state: 1    
receiver,state : 5
// 1,ready: 1  1,valid: 0  sender,state: 2    
// 1,ready: 1  1,valid: 0  sender,state: 1    
receiver,state : 6
// 1,ready: 1  1,valid: 0  sender,state: 2    
// 1,ready: 1  1,valid: 1  sender,state: 2    
// 1,ready: 1  1,valid: 1  sender,state: 1  
\end{verbatim}
\end{quote}

If the sender has not yet reached the stable state (2), the receiver will enter state 3 in an environment where the channel holds valid data, allowing it to proceed without waiting to read. The analysis provides an upper approximation of the design’s behavior, which may include scenarios that are not feasible in the actual execution. With increased parallelism, the abstract environment can grow to include many combinations of status bits and module state values. However, the deterministic nature of parallel execution in hardware reduces the likelihood of such over-approximations. As future work, we plan to explore the use of widening techniques \cite{cousot1977abstract} on state vectors to enhance performance while maintaining analytical precision.

%
%
\section{Related works}
\label{related}

The language Chisel was initially introduced by Bachrach et al. in their work on Chisel~\cite{chisel:dac2012}, and it has gained significant popularity in both hardware design and educational environments \cite{chisel:dac2012,chisel:book}.

Several authors have explored the formalization of semantics, analysis, and transformations for various hardware description languages. Thomson and Mycroft have investigated the applicability of abstract interpretation for hardware specifications written in HarPE \cite{thompson2004abstract}. 
Formalizing hardware description languages has also been explored for execution time analysis \cite{schlickling2012timing}.

CIRCT is an open-source infrastructure that provides an intermediate representation for hardware description languages, including Chisel \cite{zhao2024k}. 
As a common framework for a range of hardware description languages, it could also be used in connection with our subset of Chisel. 
The CIRCT framework contains tools for verification, symbolic evaluation, and hardware generation.

Our framework’s approach to recording intermediate stages and environment changes shares similarities with the minimal function graph interpretation of first-order functional languages \cite{jones1986data}. 
The idea of constructing data flow analysis as abstractions of formal semantics goes back to Cousot and Cousot \cite{cousot1977abstract}.

\section{Conclusion}

In this paper, we introduced a framework for the static analysis of communication within datapath architectures for FPGAs. By providing insights into communication patterns, our approach facilitates more efficient hardware design and optimization. The framework helps designers identify bottlenecks and optimize data handling, ultimately improving performance in hardware-accelerated computing. Future work will focus on extending the framework to support more complex architectures and integrating automated optimization techniques.

%
\nocite{mycroft1992minimal}

\bibliographystyle{eptcs}
\bibliography{mybibliography}

\end{document}